\documentclass{article}
\usepackage{arxiv}

\usepackage{comment}
\usepackage{graphicx}
\usepackage{color}
\usepackage{nicefrac}
\usepackage{tikz}
\usepackage{float}
\usepackage{xcolor}
\usepackage{tcolorbox}
\usepackage{caption}
\usepackage{subcaption}
\usepackage{latexsym}
\usepackage{amsthm}
\usepackage{placeins}
\usepackage{pifont}
\usepackage{siunitx}
\usepackage{booktabs}
\usepackage{makecell}
\usepackage{url}
\usepackage{algorithm}
\usepackage{algpseudocode}
\usepackage{amsmath,amssymb,amsfonts,bm,amsbsy,bbm}
\DeclareMathOperator*{\argmin}{argmin}
\usepackage[numbers]{natbib}
\usepackage[colorlinks=true,citecolor=blue]{hyperref}
\usepackage{cleveref}
\newcommand{\cmark}{\ding{51}}
\newcommand{\xmark}{\ding{55}}

\title{Equivalence Testing Under Privacy Constraints}

\author{
  Savita Pareek\thanks{Co-first author.} \\
  Sloan School of Management \\
  Massachusetts Institute of Technology \\
  \texttt{savita97@mit.edu}
  \And
  Luca Insolia\footnotemark[1] \\
  School of Pharmaceutical Sciences \\
  University of Geneva \\
  \texttt{Luca.insolia@unige.ch}
  \And
  Roberto Molinari\thanks{Corresponding author. } \\
  Department of Mathematics and Statistics \\
  Auburn University\\
  \texttt{robmolinari@auburn.edu}
  \And
  Stéphane Guerrier \\
  School of Pharmaceutical Sciences \\
  University of Geneva \\
  \texttt{Stephane.Guerrier@unige.ch}
}
\date{}
\begin{document}

\maketitle

\begin{abstract}{Protecting individual privacy is essential across research domains, from socio-economic surveys to big-tech user data. This need is particularly acute in healthcare, where analyses often involve sensitive patient information. A typical example is comparing treatment efficacy across hospitals or ensuring consistency in diagnostic laboratory calibrations, both requiring privacy-preserving statistical procedures.
However, standard equivalence testing procedures for differences in proportions or means, commonly used to assess average equivalence, can inadvertently disclose sensitive information. To address this problem, we develop differentially private equivalence testing procedures that rely on simulation-based calibration, as the finite-sample distribution is analytically intractable. Our approach introduces a unified framework, termed DP-TOST, for conducting differentially private equivalence testing of both means and proportions. Through numerical simulations and real-world applications, we demonstrate that the proposed method maintains type-I error control at the nominal level and achieves power comparable to its non-private counterpart as the privacy budget and/or sample size increases, while ensuring strong privacy guarantees. These findings establish a reliable and practical framework for privacy-preserving equivalence testing in high-stakes fields such as healthcare, among others.}
\end{abstract}

\maketitle

\renewcommand\thefootnote{}
\footnotetext{\textbf{Abbreviations:} CDF, Cumulative Distribution Function; DP, Differential Privacy; TOST, Two One-Sided Tests.}

\renewcommand\thefootnote{\fnsymbol{footnote}}
\setcounter{footnote}{1}

\section{Introduction}
Equivalence testing provides a statistical framework for demonstrating that a certain effect, such as the difference in means between two interventions or experimental outcomes, lies within a pre-specified tolerance region delimited by clinically meaningful equivalence margins \citep{schuirmann1987comparison}. 
Unlike conventional hypothesis testing to assess differences, which treats the null hypothesis as the absence of an effect, equivalence testing reverses this logic by assuming non-equivalence as the null and concluding equivalence only when enough evidence indicates otherwise.
Equivalence testing gained prominence through applications in the pharmaceutical sciences, especially for bioequivalence assessments \citep{metzler1974bioavailability,westlake1972use,westlake1976symmetrical}, which play a key role for the regulatory approval of generic drug products \citep{FDA2022,EMA2010,ICH2024,WHO2017}. 
Nowadays, equivalence testing is ubiquitous across domains, including medicine \citep{obrien2022not,wehrle2022neurodevelopmental,sansone2022tapentadol,branscheidt2022stroke}, psychology \citep{lakens2018tutorial}, economics \citep{feri2023risk}, political science \citep{aggarwal2023digital}, imaging \citep{sureshkumar2022snap}, food science \citep{meyners2012equivalence,leday2022multivariate}, engineering \citep{richter2002method,moore2022investigation}, and sports science \citep{mazzolari2022myths}. 

Parametric procedures for two-sample equivalence assessments are well-established for both binomial proportions \citep{tango1998equivalence,chen2000tests} and normal means \citep{wellek2010testing,patterson2017bioequivalence},  and they typically rely on the Two One-Sided Tests (TOST) procedure \citep{berger1996bioequivalence}. However, these classical approaches rely on unrestricted access to exact summary statistics, a requirement that poses critical privacy concerns in sensitive domains. 
In biomedical applications, for instance, sensitive patient data are increasingly exposed to potential cyberattacks on healthcare infrastructures \cite{elsayed2025cybersecurity} or the unconsented training of modern artificial intelligence tools on public data \cite{solove2024greatscrape}. Therefore, more stringent regulatory frameworks have recently been put forward to mitigate these risks, such as the Health Insurance Portability and Accountability Act (HIPAA) in the United States and the General Data Protection Regulation (GDPR) in Europe \citep{Bakare2024DATAPL}, which require moving beyond the traditional, non-private hypothesis testing framework.
Notably, privacy risks persist even when only aggregated or anonymized data are released (e.g. by linking query patterns with external sources) \cite{homer2008resolving,narayanan2008robust,gotz2011publishing}. 
To address these limitations, Differential Privacy (DP) has emerged as a leading framework to achieve formal privacy guarantees, providing individual data protection while still enabling useful insights about the population \citep{dwork2006dp,dwork2014algorithmic}. 
DP offers quantifiable guarantees by ensuring that the inclusion or exclusion of any single individual's data has a provably limited effect on the outcome of an analysis, even when arbitrary external information is available, which is typically achieved by injecting calibrated random noise into statistical functionals of interest (e.g. sample means or proportions).

In this direction, some classical inferential procedures have therefore been extended to the DP framework. 
These include confidence intervals for normal means \citep{karwa2018finite}, binomial proportions \citep{awan2020binomial}, and more general categorical data \citep{romanus2025fiducial}, as well as chi-squared tests for goodness-of-fit and independence \citep{gaboardi2016chi}.
DP procedures have also been developed to assess distributional equivalence of two discrete \citep{aliakbarpour2018dp} or continuous distributions \citep{omer2024dp_equivalence}, which resemble goodness-of-fit tests. However, to the best of our knowledge, DP tests of equivalence for a specific parameter of interest have not yet been investigated. Indeed, unlike classical hypothesis testing, equivalence testing requires precise control of rejection behavior near the boundary of the null hypothesis, making it particularly sensitive to the additional randomness introduced by DP mechanisms. In particular, standard equivalence procedures (such as TOST) rely on the joint behavior of multiple test statistics and on distributional assumptions that are generally invalidated by DP noise. As a result, naively adapting existing DP tests does not guarantee type-I error control, and valid equivalence testing under DP requires explicit recalibration of the sampling distribution to account for the privacy mechanism.

In this work, we propose a flexible simulation-based framework for  \textit{differentially private equivalence testing} through a TOST-like procedure, which is denoted as DP-TOST for brevity. Our framework applies to test statistics with known (finite or asymptotic) distributions, and we focus on two-sample equivalence assessments for binomial proportions and normal means, as they frequently arise in biomedical applications. 
Our proposal leverages the findings in Orso et al. \cite{orso2024accurate}, which offer a general strategy for accurate inference in complex settings using simulation-based methods such as indirect inference \citep{gourieroux1993indirect}. In particular, we simulate the statistics of interest from their known (asymptotic) distributions, inject the required DP noise, and employ a moment-matching procedure by comparing these simulated statistics to the empirical ones to identify the parameters that best match these two quantities. 
This matching strategy allows us to construct valid confidence intervals that can be used in a TOST procedure, ensuring DP guarantees while preserving type-I error control even under strict privacy constraints.
We demonstrate the advantages of our proposal within different simulation settings and through a real-world application using data from the ACTG175 clinical trial on HIV research \cite{hammer1996trial}, where the goal is to compare data collected by different healthcare providers or various treatment arms while ensuring that no sensitive information is disclosed. 
Moreover, we provide an implementation of the proposed DP-TOST procedures in the \texttt{cTOST} package in \texttt{R} (also available at \url{https://github.com/stephaneguerrier/cTOST}).

The remainder of this paper is structured as follows. 
\Cref{sec:intro} reviews the existing frameworks for average equivalence testing and DP.  In \Cref{sec4}, we introduce our DP-TOST proposal, describing its application to both binomial proportions and normal means. \Cref{sec5} evaluates the performance of DP-TOST through an extensive simulation study, where it is compared against its non-private counterpart. We illustrate the practical advantages of DP-TOST on the ACTG175 data in \Cref{sec6}, and conclude in \Cref{sec7} with a discussion on future research directions and implications for privacy-preserving biomedical research.

\section{Related Work}
\label{sec:intro}

\subsection{Average Equivalence Testing}\label{sec2}

Equivalence testing originated in the context of average bioequivalence \citep{westlake1972use,westlake1976symmetrical}, where the goal is to demonstrate that two formulations or treatments produce mean responses that are sufficiently similar, remaining within a pre-specified margin of clinical irrelevance. 
Bioequivalence assessments are a key element of the formal regulatory approval process for generic systemic medications \citep{FDA2022,EMA2010}, and they play a pivotal role throughout the drug development life-cycle, from pre-clinical and clinical development, to manufacturing, quality control, and pharmacovigilance \citep{patterson2017bioequivalence}. 
While average bioequivalence was originally designed for assessing differences in normal means, it naturally extends to other settings, such as differences in binomial proportions, through large-sample approximations based on asymptotic normality \citep{wellek2010testing}. 
In full generality, let the scalar parameter of interest be $\theta \in \Theta \subset \mathbb{R}$, which can represent, for example, the difference in average effects (i.e.~$\Theta = \mathbb{R}$) or proportions (i.e.~$\Theta = [-1, 1]$).
The hypotheses of interest are 
\begin{equation}\label{eq:hyp}
\text{H}_0 : \theta \in \Theta  \setminus \Theta_1 \quad \text{versus} \quad \text{H}_1 : \theta \in \Theta_1 \equiv (c_1, c_2) ,    
\end{equation}
for given constants $c_1,c_2 \in \Theta$ delimiting the equivalence region. 
Without loss of generality, for simplicity, we restrict our attention to equivalence margins symmetric around zero and set $c_0 \equiv c_2 = -c_1$.

For average equivalence, it is customary to assume that the estimator of $\theta$ follows a normal distribution $\hat{\theta} \sim \mathcal{N}(\theta, \vartheta^2)$ and an independent estimator of $\vartheta^2$ is also available and such that $v \hat{\vartheta}^2 / \vartheta^2 \sim \chi^2_{v}$, where $\chi^2_{v}$ denotes a chi-square distribution with $v$ degrees of freedom. 
For instance, in a parallel design comparing two independent groups $X_1, \ldots, X_n ~ \overset{iid}{\sim} \mathcal{N}(\mu_1, \sigma_1^2)$ and $ Y_1, \ldots, Y_m \overset{iid}{\sim} \mathcal{N}(\mu_2, \sigma_2^2)$, the parameter of interest $\theta = \mu_1 - \mu_2$ is estimated by $\hat{\theta} = \bar{X} - \bar{Y}$, where $\bar{X}$ and $\bar{Y}$ denote the sample means. 
Then, one typically employs the unpooled variance estimator of $\vartheta^2$ and uses the Welch-Satterthwaite approximation of the degrees of freedom $v$.
However, assuming that the two groups have a common variance (i.e.~$\sigma_1=\sigma_2$), the variance of $\hat{\theta}$ is estimated by the pooled variance, and the associated degrees of freedom are $v = n + m - 2$.
Similarly, considering two independent random variables $X \sim \text{binomial}(n, \pi_{1})$ and $Y \sim \text{binomial}(m, \pi_{2})$, it is of interest
to assess equivalence for the difference in proportions $\theta = \pi_{1} - \pi_{2}$. Therefore, we have that $\hat{\theta} = \bar{X} -\bar{Y}$, where $\bar{X}$ and $\bar{Y}$ denote the observed proportion of successes in the two groups.
To estimate its variance $\vartheta^2$ without assuming equal proportions (and thus avoiding the bias of pooled estimators in equivalence settings), one often relies on the unpooled variance estimator \citep{barker2001EquivalenceTesting}.

The TOST procedure \cite{schuirmann1987comparison} represents the most established framework to assess the hypotheses in \eqref{eq:hyp}.
It decomposes such composite hypotheses into two one-sided tests of hypotheses: H$_{0}^{(L)}: \theta \leq -c_0$ versus H$_{1}^{(L)}: \theta > -c_0$, and H$_{0}^{(U)}: \theta \geq c_0$ versus H$_{1}^{(U)}: \theta < c_0$. 
Hence, the TOST rejects $\text{H}_{0}\equiv\text{H}_{0}^{(L)} \cup \, \text{H}_{0}^{(U)}$ in favor of $\text{H}_{1}\equiv\text{H}_{1}^{(L)} \cap \, \text{H}_{1}^{(U)}$ at the nominal significance level $\alpha  \in (0, 1/2]$ when both marginal hypotheses H$_{0}^{(L)}$ and H$_{0}^{(U)}$ are rejected.
Specifically, regardless of the specific study design (e.g. parallel, paired or crossover), the TOST procedure relies on two test statistics:  
$
T_L \equiv (\hat{\theta}+c_0) / \hat{\vartheta}
$
and 
$
T_U \equiv (\hat{\theta}-c_0) / \hat{\vartheta},
$
to respectively test for $\text{H}_{0}^{(L)}$ and $\text{H}_{0}^{(U)}$. 
Thus, at a significance level $\alpha$, the TOST rejects
$\text{H}_{0}$ in \eqref{eq:hyp} if both tests simultaneously reject their marginal null hypotheses, that is, when $T_L \geq q_{\alpha}$ and $T_U \leq q_{1-\alpha}$,
where $q_{\alpha}$ denotes the $\alpha$ quantile of the (asymptotic) reference distribution (e.g. a
$t$-distribution with $v$ degrees of freedom for normal means, and a standard normal for binomial proportions). 
Since both tests (asymptotically) control the type-I error rate at the nominal level $\alpha$, the intersection-union principle ensures that the resulting test for \eqref{eq:hyp} is performed at the level $\alpha$ \citep{berger1996bioequivalence}.
Operationally, based on the interval-inclusion principle \citep{berger1996bioequivalence}, the TOST decision is equivalent to the test procedure that leads to a declaration of equivalence when the $1-2\alpha$  confidence interval for  $\theta$, say $\text{CI}(\theta, \alpha) \equiv [\hat{\theta} + q_{\alpha} \hat{\vartheta} , \hat{\theta} + q_{1-\alpha} \hat{\vartheta}]$, falls entirely within the equivalence margins $(-c_0, c_0)$.

\subsection{Differential Privacy (DP)}\label{sec3}

DP provides a rigorous and quantifiable framework for protecting individual-level information, such as medical diagnoses from health records, income data from census surveys, or user preferences from social media activity, during statistical analysis. As noted by Dwork\citep{dwork2014algorithmic}:
\begin{quote}
\textit{“Differential privacy describes a promise, made by a data holder, or curator, to a data subject: ‘You will not be affected, adversely or otherwise, by allowing your data to be used in any study or analysis, no matter what other studies, data sets, or information sources are available.’”}
\end{quote}
More specifically, while various forms of DP exist, in this work we consider the most common form (called \textit{central} DP), which assumes that there is a trusted curator who has access to all the raw data (e.g. a hospital data center). In this setting, there may be internal or external analysts who are not granted direct access to the data and therefore submit queries to the curator in order to obtain summary statistics, such as sample means or proportions. However, the release of these statistics can still violate patient privacy, since a malicious attacker (through linkage attacks or other strategies) can issue a sequence of queries that enables reconstruction of individual-level information. Indeed, Dinur et al.\citep{dinur2003revealing} showed that an adversary issuing sufficiently many aggregate queries can accurately reconstruct sensitive records when the responses are not adequately protected. To address this risk, DP places an explicit limit on the amount of information that can be released (informally referred to as the ``\textit{privacy budget}''), and enforces this limit by adding calibrated noise to the output of each query. As a consequence, the added noise guarantees a form of plausible deniability for any individual or patient regarding whether or not they were included in the dataset.

With the above in mind, we now formally define the DP framework. Let \( \mathcal{M} : \mathcal{X}^n \rightarrow \mathcal{R} \) be a randomized mechanism applied to a dataset \( \mathbf{X} \in \mathcal{X}^n \), where \( \mathcal{X}^n \) denotes the space of databases with \( n \) rows. We say that two databases \( \mathbf{X}, \mathbf{X}' \in \mathcal{X}^n \) are \textit{neighboring} if they differ in exactly one row, representing the worst-case scenario in which an attacker has access to all but one individual’s data and seeks to determine whether that individual is included in the database. Using the definition of \textit{pure} DP and denoting the privacy budget by $\epsilon$, a mechanism \( \mathcal{M} \) satisfies \(\epsilon\)-DP if, for all neighboring databases \( \mathbf{X} \) and \( \mathbf{X}' \), and for all measurable subsets \( S \subset \mathcal{R} \), we have
\begin{equation}
\Pr[\mathcal{M}(\mathbf{X}) \in S] \leq e^{\epsilon}  \Pr[\mathcal{M}(\mathbf{X}') \in S].
\label{eq:dp}
\end{equation}
In lay terms, the probability of any given outcome of the DP mechanism is nearly the same regardless of whether an individual is included in the database or not (with similarity being inversely proportional to the privacy budget $\epsilon$). This definition, introduced by Dwork\citep{dwork2006calibrating}, guarantees that the inclusion or exclusion of a single individual has a bounded influence on the distribution of the output, thereby preserving individual-level privacy. The strength of this guarantee depends on the privacy budget $\epsilon$: smaller values provide stronger privacy protection (at the cost of increased noise), while larger values yield weaker privacy guarantees but more accurate outputs. Typical values of $\epsilon$ are often reported in the range \( [0.1, 1] \), although substantially larger values have been used in practice (for example, in U.S. Census applications)~\citep{alabi2022hypothesis}.

To satisfy the definition in \eqref{eq:dp}, random noise must be injected into the response to each query. The scale of this noise is determined by the \emph{sensitivity} of the function \( f \) of interest (e.g. a mean or proportion). While different notions of sensitivity exist, in this work we focus on the \textit{global sensitivity}, defined as
\[
\Delta f \equiv \max_{\mathbf{X},\mathbf{X}'} \|f(\mathbf{X}) - f(\mathbf{X}')\|,
\]
where the maximum is taken over all neighboring databases and $\|\cdot\|$ denotes a norm (typically the $\ell_2$-norm). This quantity captures the maximum change in the query output attributable to a single individual. The core idea of DP is therefore to inject enough noise to mask this worst-case change, for instance when an individual with extreme values is replaced by another individual with equally extreme values in the opposite direction. While some queries have well-defined global sensitivities (for example, proportions satisfy $\Delta f = 1/n$), others do not. In such cases, the curator must impose \textit{a priori} bounds on the data, say $[a, b]$, which are then used to ``\textit{clamp}'' observations (i.e. replacing values outside the chosen range with the nearest boundary value) in order to define a finite global sensitivity, as is commonly done for sample means for example.

Building on these concepts, the DP literature has introduced a variety of mechanisms that satisfy the definition in \eqref{eq:dp} for different statistical tasks. Representing the query/statistics of interest as $f$, the mechanism considered in this work is the most widely used DP approach for numerical queries, namely the \textit{additive} mechanism~\citep{dwork2006calibrating}. This mechanism randomizes the function \( f : \mathcal{X}^n \rightarrow \mathbb{R} \) by adding noise scaled to the function’s global sensitivity. Specifically, the mechanism is defined as
\begin{equation}
\label{eq:dp_mech}
    \mathcal{M}(\mathbf{X}) \equiv f(\mathbf{X}) + U,
\end{equation}
where $U$ is a zero-mean random variable whose scale is proportional to $\Delta f / \epsilon$. For example, when $U$ follows a Laplace distribution with scale parameter $\Delta f / \epsilon$, the resulting mechanism satisfies \(\epsilon\)-DP by construction \citep{dwork2014algorithmic}. While the remainder of this paper focuses on the additive Laplace mechanism as a running example, the proposed methodology naturally extends to a broader class of additive mechanisms based on symmetric, zero-mean noise distributions, including the Gaussian mechanism.

\section{Differentially Private TOST} \label{sec4}

This section presents a unified framework for constructing DP confidence intervals and equivalence tests via a simulated moment-matching approach.  This formulation supports valid inference for a wide range of outcomes, including rates, probabilities, and bounded clinical scores, commonly encountered in biomedical and public health applications. In particular, we employ the technique put forward and studied in Orso et al. \citep{orso2024accurate} 
(see also Gourieroux, Guerrier et al.\citep{gourieroux1993indirect,Guerrier2019bias} and references therein).
More in detail, supposing for simplicity that we are interested in a general parameter $\bm{\theta} \in \bm{\Theta} \subset \mathbb{R}^p$ for which we have an estimator $\hat{\bm{\eta}}$ (we refer to this as an auxiliary estimator) which may be biased and/or even inconsistent with respect to the true parameter (which we denote as $\bm{\theta}_0$). In this case, under certain regularity conditions, a new moment-matching (minimum distance) estimator can be derived with good statistical properties as follows:
\begin{equation}
\label{eq:ib}
    \hat{\bm{\theta}} \in \hat{\bm{\Theta}} \equiv \underset{\bm{\theta} \in \bm{\Theta}}{\argmin} \, \|\hat{\bm{\eta}}(\bm{\theta}_0) - \hat{\bm{\eta}}^*(\bm{\theta})\|,
\end{equation}
where $\hat{\bm{\eta}}(\bm{\theta}_0)$ represents the auxiliary estimator from the observed sample (generated from the unobserved $\bm{\theta}_0$) and $\hat{\bm{\eta}}^*(\bm{\theta})$ represents the auxiliary estimator computed on a simulated sample generated from a generic parameter value $\bm{\theta}$. The latter estimator will therefore depend on a simulated sample (say $X^*$) whose randomness is determined by a seed value (say $\omega^*$) and for which one can then find a value of $\bm{\theta}$ that solves the problem in \eqref{eq:ib}. If one generates multiple seeds (say $\omega^*_1, \ldots, \omega^*_H$ with $H \gg 0$) and consequently multiple samples, Orso et al. \citep{orso2024accurate} show that the sequence of estimators $\{\hat{\bm{\theta}}^h\}_{h=1}^H$ can be used to perform inference using a percentile-based approach on this distribution. In a similar manner, if obtaining a statistic $\hat{\nu} \equiv \varphi(\hat{\bm{\theta}})$, where $\varphi: \mathbb{R}^p \to \mathbb{R}$ is a continuous function, then the sequence $\{\hat{\nu}^h\}_{h=1}^H$, where $\hat{\nu}^h \equiv \varphi(\hat{\bm{\theta}}^h)$, can also be used to perform inference on the true function $\nu_0 \equiv \varphi(\bm{\theta}_0)$.

In this work our auxiliary estimator $\hat{\bm{\eta}}$ is the privatized (DP) estimator defined in \eqref{eq:dp_mech} where the function $f$ is the estimator of the sample proportion or sample mean (with clamped values) and the noise is calibrated to the sensitivity $\Delta f$ of each of them respectively. As a result, since we are testing equivalence between two statistics, we will focus on the distribution of the parameter vector $\hat{\bm{\theta}} \equiv [\hat{\theta}_1, \hat{\theta}_2]^T$,  where $\hat{\theta}_i \in \mathbb{R}$ is the estimator in \eqref{eq:ib} computed on sample $i$. More specifically, we will focus on the derived statistic $\hat{\nu} \equiv \hat{\theta}_1 - \hat{\theta}_2$ which constitutes a continuous function of $\hat{\bm{\theta}}$. Following this method we can therefore compute $H$ solutions for the latter quantity to obtain the sequence $\{\hat{\nu}^h\}_{h=1}^H$ based on which we can define the $\delta$-quantile of the estimator $\hat{\nu}$ as:
\begin{equation*}
    \hat{\nu}_{\delta} \equiv \inf \left\{ r \in \mathbb{R} \; : \; \Pr\left( \hat{\nu} \le r \,\middle|\, \hat{\bm{\eta}} \right) \ge \delta \right\}.
\end{equation*} 
This definition therefore allows us to build the $1 - 2\alpha$ (asymptotic) confidence interval using the percentile method as follows:
\begin{equation}
    \label{eq.qci}
    \text{CI}(\nu_0, \alpha) \equiv [\hat{\nu}_{\alpha} \, , \, \hat{\nu}_{1 - \alpha}] \,.
\end{equation}
Consequently, based on the interval-inclusion principle, equivalence is declared when $\text{CI}(\nu_0, \alpha) \subset (-c_0, c_0)$.

In this perspective, we firstly adapt this approach to assessing the equivalence between two population proportions and then extend it to general mean parameters with pre-defined bounds for the data over a fixed interval $[a,b]$ (potentially group-specific), the latter being imposed to define global sensitivity and enable structured noise calibration under DP. In particular, among others, the use of the above framework allows to easily address more challenging settings including possible different sample sizes between groups. Hereinafter, we refer to this proposed framework, which is based on \eqref{eq:ib}, as the \textit{DP-TOST} approach.

\subsection{DP-TOST for Proportions}

To present the case for equivalence of proportions, let us start by focusing on one binary-valued sample that we denote as \( X_1, \dots, X_{n} \overset{iid}{\sim} \mathrm{Bernoulli}(\pi_{1}) \), after which we will then introduce the second binary-valued sample \( Y_1, \dots, Y_m \overset{iid}{\sim} \mathrm{Bernoulli}(\pi_{2}) \) to perform the two-sample test. With this in mind, the sample proportion is given by
\[
\bar{X} = \frac{1}{n} \sum_{i=1}^n X_i.
\]
To ensure DP we release adequately privatized versions of this statistic through the following additive mechanism:
\[
\hat{p}_1 \equiv \bar{X} + U_1,
\]
where $U_1$ is an independent zero-mean random variable with variance proportional to the global sensitivity of the statistic.  As mentioned in \Cref{sec3}, in this work we will consider \( U_1 \sim \mathrm{Laplace}(0, \lambda_1) \) which guarantees pure $\epsilon$-DP and where the noise scale \( \lambda_1 = (n\epsilon)^{-1} \) is based on the global sensitivity of sample proportion (i.e.  $\Delta f = n^{-1}$). We now have that $\hat{p}_1$ plays the role of $\hat{\eta}(\theta_0)$ in \eqref{eq:ib} where $\hat{p}_1 = \bar{X}(\pi_{1}) + U_1$ is the DP proportion computed on the observed sample, with underlying non-private sample proportion $\bar{X}(\pi_{1})$ (and with $\pi_{1}$ playing the role of $\theta_0$). Following this, using $\|\cdot\|$ to denote the $l_2$-norm, we would therefore aim to solve the following matching problem:
$$\check{p}_1 \equiv \underset{\pi \in [0,1]}{\argmin} \, \|\hat{p}_1 - \hat{p}_1^*(\pi)\|,$$
where $\hat{p}_1^*(\pi) \equiv \bar{X}(\pi) + U_1^*$ and $U_1^*$ is independent from and follows the same distribution as $U_1$. Given the relatively simple additive structure, we could decide to solve the matching explicitly by finding the value of $\pi$ such that $\hat{p}_1 = \hat{p}_1^*(\pi)$. Indeed, expanding the term on the right of the latter equality and rearranging would lead to the solution:
$$\check{p}_1 = F^{\, -1}(\hat{p}_1 - U_1^*),$$
where $F$ is the Cumulative Distribution Function (CDF) of a binomial random variable. The problem here is that inverting a discrete CDF gives interval solutions, therefore we use an approximation to find a point solution. More specifically, using the central limit theorem, the privatized estimators approximately follow:
\[
\hat{p}_1^*(\pi) \overset{d}{=} \pi + \sqrt{\frac{\pi(1 - \pi)}{n}} Z_1^*  + U_1^* + o_p(1), 
\]
where \( Z_1^* \sim \mathcal{N}(0,1) \) is an independent standard normal random variable. Hence we can now represent $\hat{p}_1^*(\pi)$ through the above approximation and solve the matching explicitly. More in detail, by rewriting the matching problem we want to find the minimum in $\pi$ of the following least-squares problem:
\begin{equation} \label{eq:obj_fun}
\check{\pi}_1 \equiv \underset{\pi \in [0,1]}{\argmin} \,\Bigg\| \hat{p}_1 - \pi - \sqrt{\frac{\pi(1 - \pi)}{n}} Z_1^* - U_1^* \Bigg\|.
\end{equation}
Since we now have an explicit (approximate) form for $\hat{p}_1(\pi)$, we can rewrite this as a quadratic equation and also find explicit solutions for $\pi$. To do so, we define the following quantities
\[
\delta \equiv \frac{Z_1^*}{\sqrt{n}}, \quad \gamma \equiv \delta^2, \quad
\Lambda \equiv -4\hat{p}_1^2 + 4\hat{p}_1 + \gamma + 8\hat{p}_1 U_1^* - 4(U_1^*)^2 - 4U_1^*.
\]
With these, if \( \Lambda \geq 0 \), the two candidate roots for the matching problem are given by:
\begin{equation}\label{eq:candi}
\begin{aligned}
\check{\pi}_1^{(l)} &\equiv \frac{-\delta \sqrt{\Lambda} + (2\hat{p}_1 + \gamma - 2U_1^*)}{2(\gamma + 1)}, \\
\check{\pi}_1^{(r)} &\equiv \frac{\delta \sqrt{\Lambda} + (2\hat{p}_1 + \gamma - 2U_1^*)}{2(\gamma + 1)}.
\end{aligned}
\end{equation}
We restrict our search to roots that lie in the interval $[0,1]$, and if they are not found the procedure is repeated up to a certain maximum number of attempts. Among two valid roots, we evaluate the objective in \eqref{eq:obj_fun} and select the solution with the smallest loss, ensuring that the corresponding $\check{\pi}_1 \in \{ \check{\pi}_1^{(l)}, \check{\pi}_1^{(r)} \}$ allows to best match $\hat{p}_1^*(\check{\pi}_1)$ to the observed privatized estimator $\hat{p}_1$. 
By repeating this procedure \( H \) times, where each time we produce new independent draws of the variables $Z_1^*$ and $U_1^*$, we obtain a collection of reconstructed values $\{ \check{\pi}_1^{h}\}_{h = 1}^H$ which can be used for inference on $\pi_{1}$. All of the above procedure can be applied in the exact same manner for the second sample $\{Y_1, \hdots, Y_m\}$, using $U_2^* \sim \mathrm{Laplace}(0, \lambda_2)$ with $ \lambda_2 = (m\epsilon)^{-1} $, to obtain $\{ \check{\pi}_2^{h}\}_{h = 1}^H$. As a result, each pair $(\check{\pi}_1^{h}, \check{\pi}_2^{h})$ can be used to approximate the sampling distribution of \( \hat{\nu} = \check{\pi}_1 - \check{\pi}_2 \) by defining $\hat{\nu}^h = \check{\pi}_1^h - \check{\pi}_2^h$ and using the sequence $\{\hat{\nu}^{h}\}_{h=1}^H$ as the approximate distribution of $\hat{\nu}$. The steps involved in generating this distribution are outlined in Algorithm~\ref{alg:dp_tost_prop}, as well as the percentile approach to construct CIs and assess equivalence.

\begin{algorithm}
\caption{DP-TOST for binomial proportions}
\label{alg:dp_tost_prop}
\begin{algorithmic}[1]
\State \textbf{Input:} Privatized estimates $\hat{p}_1, \hat{p}_2$, sample sizes $n, m$, privacy budget $\epsilon$, number of Monte Carlo replicates $H$, significance level $\alpha$, equivalence margin $c_0$
\State \textbf{Output:} $(1-2\alpha)$ confidence interval for $\pi_1-\pi_2$ and the equivalence decision
\For{$h = 1, \dots, H$}
    \State Generate independent $(Z_1^{*h}, Z_2^{*h}) \overset{iid}{\sim} \mathcal{N}(0,1)$
    \State Generate independent $U_1^{*h} \overset{iid}{\sim} \mathrm{Laplace}\{0, (n \epsilon)^{-1}\}$ and $U_2^{*h} \overset{iid}{\sim} \mathrm{Laplace}\{0, (m \epsilon)^{-1}\}$
    \State Obtain candidate solutions $(\check{\pi}_1^{h}, \check{\pi}_2^{h})$ using \cref{eq:candi}
    \State If solutions are outside $[0,1]$, discard and repeat from Step 4
    \State Compute $\hat{\nu}^{h} = \check{\pi}_1^{h} - \check{\pi}_2^{h}$
\EndFor
\State Compute the $1-2\alpha$ confidence interval $\mathrm{CI}(\pi_1-\pi_2, \alpha)$  using eq. \eqref{eq.qci}
\State Set $\mathrm{Decision} \leftarrow \mathbbm{1}\{ \mathrm{CI}(\pi_1-\pi_2, \alpha) \subset (-c_0, c_0) \}$, where $\mathbbm{1}\{\cdot\}$ denotes the indicator function
\State \Return $\mathrm{CI}(\pi_1-\pi_2, \alpha)$ and $\mathrm{Decision}$
\end{algorithmic}
\end{algorithm}

\subsection{DP-TOST for Means }
\label{sec.means}
We extend our framework to construct DP equivalence tests for two population means, where the outcome is assumed to be real-valued. In this setting, to guarantee privacy a common approach is to clamp the data within pre-specified bounds in order to define global sensitivity and calibrate the magnitude of noise required for DP \citep{dwork2014algorithmic}. More specifically, let \( X_1, \dots, X_n \overset{iid}{\sim} \mathcal{N}(\mu_1, \sigma_1^2) \) and \( Y_1, \dots, Y_m \overset{iid}{\sim} \mathcal{N}(\mu_2, \sigma_2^2) \) be independent random variables: since the support of Gaussian variables is unbounded, the standard approach to achieve DP is to assume that these variables can be reasonably clamped within \textit{a priori} known bounds, i.e. \( X_i \in [a_1, b_1] \) and \( Y_j \in [a_2, b_2] \) with constants $a_k, b_k \in \mathbb{R}$ for $k=1,2$ and for all $i,j$. Hence, let us denote the clamped samples as $ \tilde{X}_i \equiv \min\{\max(X_i, a_1), b_1\}$ and $ \tilde{Y}_j \equiv \min \{ \max(Y_j, a_2), b_2\} $, for $i=1,\ldots,n$ and $j=1,\ldots,m$.
We then define:
\[
\mathbf{T}_X \equiv (m_X, s_X)^T, \quad \mathbf{T}_Y \equiv (m_Y, s_Y)^T,
\]
where \( m_X \) and \( m_Y \) are the sample means, and \( s_X \), \( s_Y \) are the sample standard deviations of the clamped data. To achieve DP, calibrated noise is independently added to both the sample mean and standard deviation. The noise scale is determined by the global sensitivity of each statistic, which depends on the range \([a_k, b_k]\) and the sample size. The privatized summary statistics are:
\[
\hat{\mathbf{T}}_X \equiv (m_X + U_1, \,s_X + U_2)^T, \quad \hat{\mathbf{T}}_Y \equiv (m_Y + U_3, \,s_Y + U_4)^T,
\]
where \( U_1, \dots, U_4\) are independent zero-mean random variables representing the noise to guarantee DP, with each noise term calibrated using a scaled additive mechanism (in this work we focus on Laplacian noise as mentioned in \Cref{sec3}). Since for this work we consider pure \(\epsilon\)-DP, to ensure that the joint release satisfies this requirement (and since the samples are independent) the privacy budget is divided equally between the mean and the standard deviation for each sample. Hence, the scale parameter for the privacy noise $U$ is given by
\[
\tau_i = \frac{\Delta f_i}{\epsilon/2},
\]
where \(\epsilon > 0\) denotes the total privacy budget for the DP equivalence test and \(\Delta f_i\) is the global sensitivity of a statistic (in this case mean or standard deviation) for sample $i$. Generally, for data bounded in \([a, b]\), the global sensitivities for sample mean and standard deviation are respectively
\[
\Delta f_{\text{mean}} \equiv \frac{b-a}{n} 
\qquad \text{and} \qquad 
\Delta f_{\text{sd}} \equiv \frac{b-a}{\sqrt{n-1}}.
\]
This calibration guarantees that the privatized statistics \(\hat{\mathbf{T}}_X\) and \(\hat{\mathbf{T}}_Y\) together satisfy \(\epsilon\)-DP (other forms of DP can be guaranteed in a similar manner). Now, to recover estimates of the underlying population means from these privatized summaries, taking the first group $\tilde{X}_1, \ldots, \tilde{X}_n$ as a reference, we solve the following matching problem:
\begin{equation}\label{eq:obj1}
\check{\mathbf{T}}_X \equiv \argmin_{\mu, \sigma} \left\| \hat{\mathbf{T}}_X - \hat{\mathbf{T}}_X^*\left(\mu, \sigma\right) \right\|,
\end{equation}
where $\hat{\mathbf{T}}_X^*\left(\mu, \sigma\right)$ is the privatized simulated parameter vector defined as
$$
\hat{\mathbf{T}}_X^*\left(\mu, \sigma\right) \equiv \left[ m_X^*(\mu, \sigma) - U_1^*, \ \mathrm{s}_X^*(\mu, \sigma) - U_2^* \right]^T,
$$
where $m_X^*(\mu, \sigma)$ and $\mathrm{s}_X^*(\mu, \sigma)$ are respectively the simulated sample mean and standard deviation implied by the parameters $\mu$ and $\sigma$, i.e. computed on an independent simulated sample $\tilde{X}_i^* = \min\{\max(X_i^*, a_1), b_1\}$, where $X_i^*=\mu + \sigma Z_1^*$ and $Z_1^*\sim\mathcal{N}(0,1)$. Moreover, $U_1^*$ and $U_2^*$ are the simulated independent noise respectively added to the sample mean and standard deviation to guarantee DP. Following the same logic for the second group $\tilde{Y}_1, \ldots, \tilde{Y}_m$, we obtain $\check{\mathbf{T}}_Y$, based on $m_Y^*(\mu, \sigma)$ and $\mathrm{s}_Y^*(\mu, \sigma)$ representing the simulated mean and standard deviation under parameters $\mu$ and $\sigma$ and independent realizations of $Z_2^*\sim\mathcal{N}(0,1)$, with $U_3^*$ and $U_4^*$ denoting the corresponding privatization noise.

Unlike the case for proportions, and also due to the clamping of the data, no closed-form solutions are available when both the mean and standard deviation are privatized. For this problem therefore, we employ numerical optimization to recover parameter estimates under privacy constraints for both groups. Since our interest lies in the means of the samples (while the standard deviations are nuisance parameters that nevertheless need to be appropriately estimated for inference), we focus on the difference in the means which are represented by the first elements in the vectors $\check{\mathbf{T}}_X$ and $\check{\mathbf{T}}_Y$. By denoting these estimated means as $\check{\mu}_1$ and $\check{\mu}_2$ respectively, our final statistic of interest is therefore $\hat{\nu} = \check{\mu}_1 - \check{\mu}_2$ whose distribution is estimated by $\{\hat{\nu}\}_{h=1}^H$, where $\hat{\nu}_h = \check{\mu}_1^h - \check{\mu}_2^h$. The procedure for constructing differentially private confidence intervals and assessing equivalence for normal means is outlined in Algorithm~\ref{alg:dp_tost_mean}.

\begin{algorithm}
\caption{DP-TOST for normal means}
\label{alg:dp_tost_mean}
\begin{algorithmic}[1]
\State \textbf{Input:} Privatized estimates $\hat{\mathbf{T}}_X$, $\hat{\mathbf{T}}_Y$, sample sizes $n, m$, data clamping bounds $[a_1, b_1]$, $[a_2, b_2]$, privacy budget $\epsilon$, number of Monte Carlo replicates $H$, significance level $\alpha$, equivalence margin $c_0$
\State \textbf{Output:} $(1-2\alpha)$ confidence interval for $\mu_1-\mu_2$ and the equivalence decision
\For{$h = 1, \dots, H$}
     \State Generate independent $(Z_1^{*h}, Z_2^{*h}) \overset{iid}{\sim} \mathcal{N}(0,1)$ 
    \State Generate independent 
    $U^{*h}_{1} \overset{iid}{\sim}
    \text{Laplace}\!\left(0, \tfrac{b_{1} - a_{1}}{n(\epsilon/2)}\right)$ and 
    $U^{*h}_{2} \overset{iid}{\sim}
    \text{Laplace}\!\left(0, \tfrac{b_{1} - a_{1}}{\sqrt{n-1}(\epsilon/2)}\right)$
    \State Generate independent 
    $U^{*h}_{3} \overset{iid}{\sim}
    \text{Laplace}\!\left(0, \tfrac{b_{2} - a_{2}}{m(\epsilon/2)}\right)$ and 
    $U^{*h}_{4} \overset{iid}{\sim}
    \text{Laplace}\!\left(0, \tfrac{b_{2} - a_{2}}{\sqrt{m-1}(\epsilon/2)}\right)$
    \State Obtain candidate solutions $(\check{\mathbf{T}}_X^{h},\check{\mathbf{T}}_Y^{h})$ using \cref{eq:obj1}
    \State If solutions for group $k$ are outside $[a_k,b_k]$, discard and repeat from Step 4
    \State Compute \( \hat{\nu}^{h} = \check{\mu}_{1}^{h} - \check{\mu}_{2}^{h}\)
\EndFor
\State Compute the $1-2\alpha$ confidence interval $\mathrm{CI}(\mu_1-\mu_2, \alpha)$  using eq. \eqref{eq.qci}
\State Set $\mathrm{Decision} \leftarrow \mathbbm{1}\{ \mathrm{CI}(\mu_1-\mu_2, \alpha) \subset (-c_0, c_0) \}$, where $\mathbbm{1}\{\cdot\}$ denotes the indicator function
\State \Return $\mathrm{CI}(\mu_1-\mu_2, \alpha)$ and $\mathrm{Decision}$
\end{algorithmic}
\end{algorithm}

\section{Simulation Study}
\label{sec5}

In this section, we evaluate the empirical performance of the proposed DP-TOST framework by presenting simulation results for two-sample equivalence tests on both proportions and means. 
We assess type-I error rates and statistical power across a variety of scenarios, which are summarized in \Cref{tab:sim_table}. Specifically, we consider different privacy budgets ($\epsilon$), sample sizes (keeping $n=m$), and parameter values (true proportions or means), using the traditional non-private TOST procedure as a benchmark. 
For tests on means, we also consider different data clamping schemes (using the same bounds across the two groups), highlighting how this choice affects the resulting test procedure. 
We fix the nominal significance level at $\alpha=5\%$ and the number of solutions used for the DP-TOST is fixed at $H = 10^3$. Moreover each simulation scenario is evaluated through $B=10^4$ Monte Carlo replications. Results for one-sample equivalence assessments, unequal sample sizes or group-specific clamping bounds provide similar patterns and are not reported for brevity.

\begin{table}[h]
    \centering
    \includegraphics[width=1\textwidth]{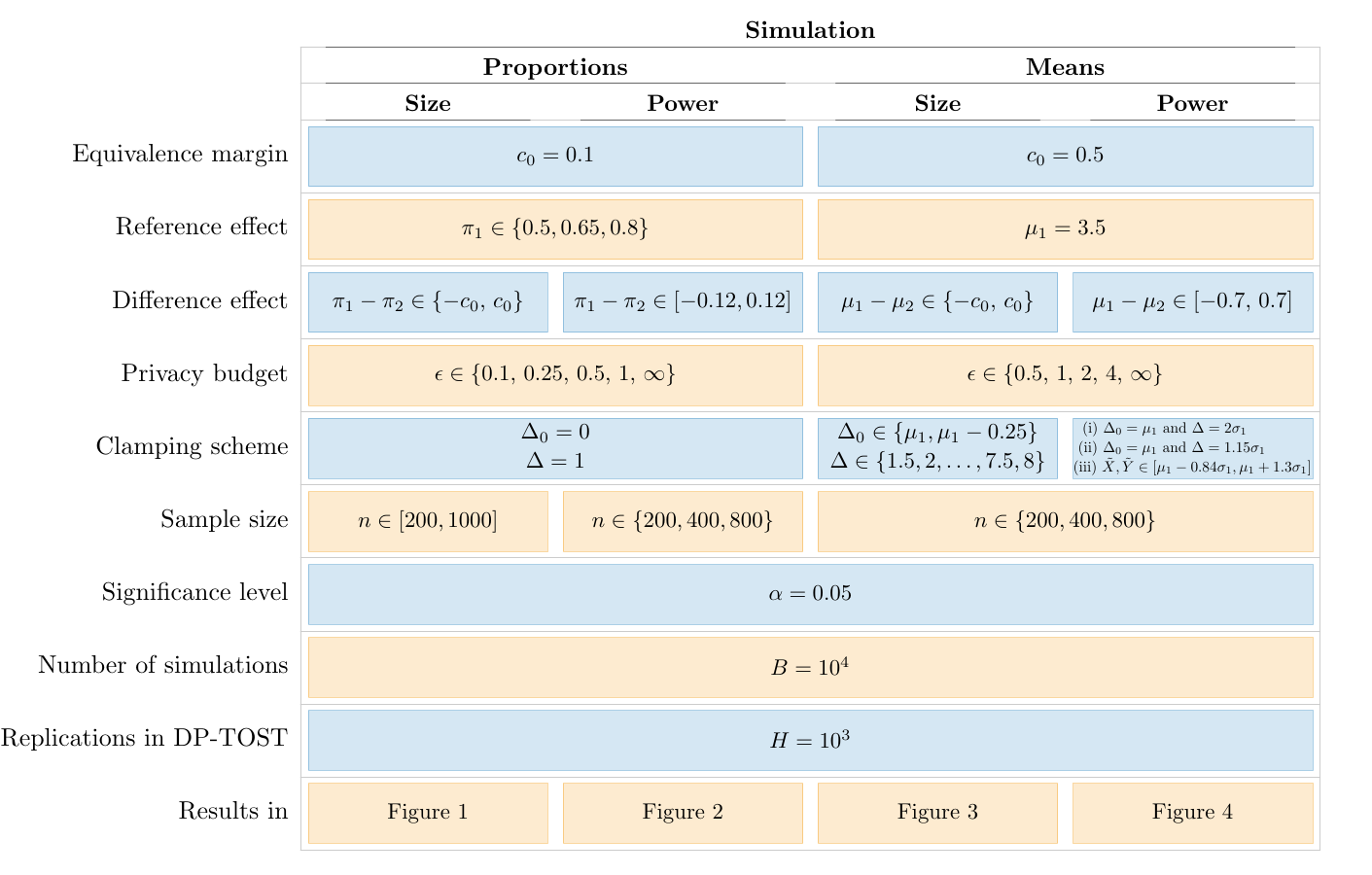}
    \caption{Simulation setting for two-sample DP-TOST.}
    \label{tab:sim_table}
\end{table}

\subsection{Proportions}

\begin{figure}[h]
    \centering
    \includegraphics[width=0.85\textwidth]{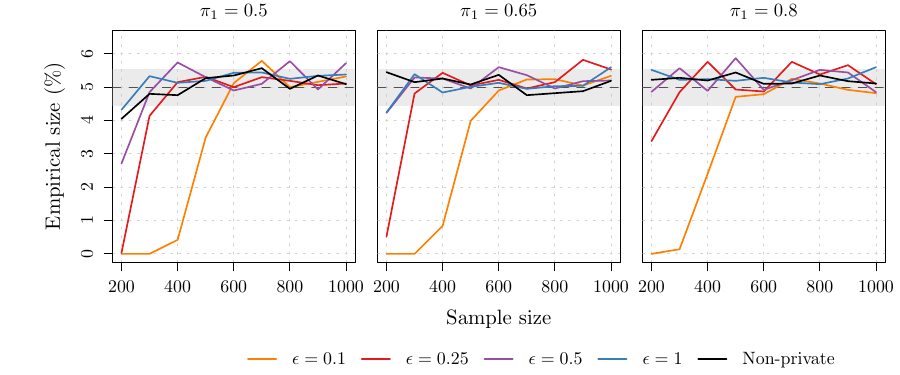}
    \caption{Empirical test size (y-axis) as a function of the sample size (x-axis) for the non-private TOST and DP-TOST with different privacy budgets $\epsilon$ when assessing equivalence for the difference of two proportions. Each panel corresponds to a fixed value of $\pi_1 \in \{0.5, 0.65, 0.8\}$, and the gray shaded region denotes the 95\% simulation error around the nominal significance level $\alpha=5\%$.}
    \label{fig:level_prop}
\end{figure}

We first investigate the finite-sample properties of the DP-TOST framework when assessing equivalence for two proportions. We set the reference proportion to $\pi_1 \in \{0.5,0.65,0.8\}$, with a fixed equivalence margin $c_0=0.1$.
\Cref{fig:level_prop} illustrates the empirical test size (y-axis), corresponding to the maximum probability of rejecting H$_0$ when $ \pi_1 - \pi_2 \in \{- c_0, c_0$\}, as a function of the sample size $n$ (x-axis) across the considered $\pi_1$'s.
These results show that DP-TOST effectively maintains type-I error control across all considered privacy budgets. 
In particular, for smaller sample sizes ($n< 400$) we observe that looser privacy budgets (e.g. $\epsilon \geq 0.5$) lead to test procedures that closely mimic the behavior of the non-private TOST, while stricter privacy budgets ($\epsilon \leq 0.25$) can lead to conservative test procedures.
This conservativeness is partly due to the addition of large DP noise, which increases the variance of the test statistic, combined with the normal approximation used in small samples. However, as the sample size increases ($n\geq500$), the performance of DP-TOST converges rapidly to that of a non-private TOST across all considered privacy budgets, and the type I error remains very close to the nominal significance level $\alpha$ (generally remaining within a 95\% simulation error tolerance displayed as a gray shaded region around $\alpha$). 

\begin{figure}[h]
    \centering
    \includegraphics[width=0.67\textwidth]{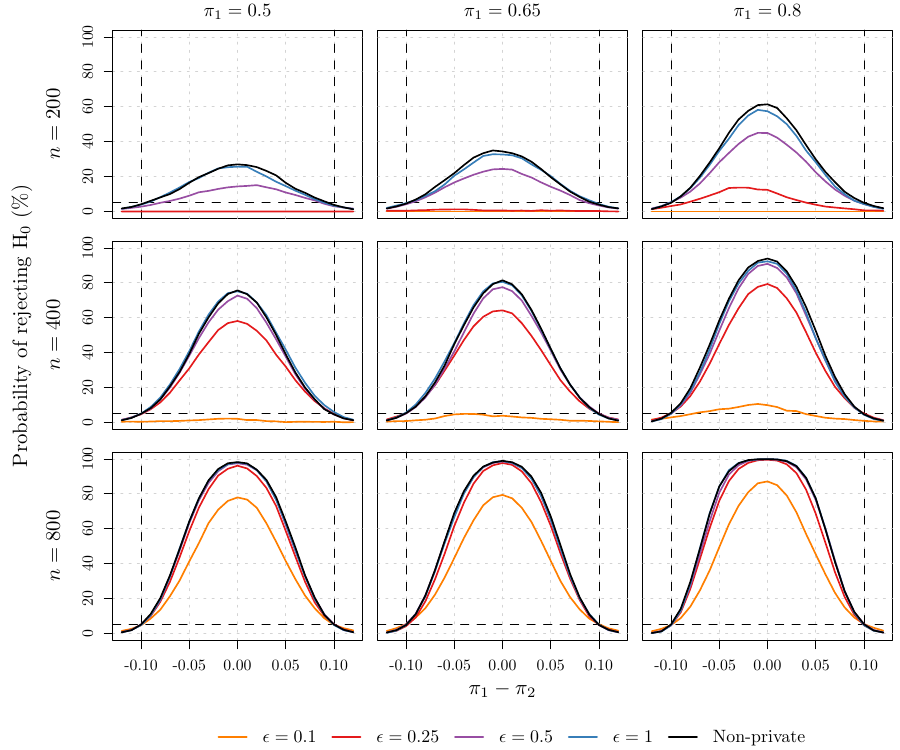}
    \caption{Empirical probability of rejecting H$_0$ (y-axis) as a function of the true difference in proportions (x-axis) for the non-private TOST and DP-TOST with different privacy budgets $\epsilon$. Each column corresponds to a fixed value of $\pi_1 \in \{0.5, 0.65, 0.8\}$ and each row represents a given sample size $n \in \{ 200, 400, 800\}$. Vertical dashed lines denote the equivalence margins $c_0=0.1$, and the horizontal dashed line represents the nominal significance level $\alpha=5\%$.}
   \label{fig:power_prop}
\end{figure}

More generally, \Cref{fig:power_prop} shows the probability of rejecting H$_0$ (y-axis) as a function of the true difference in proportions $\pi_1 - \pi_2 \in [-0.12, 0.12]$ (x-axis), when varying the sample size $n \in \{ 200, 400, 800\}$ (across the rows) and the reference proportion $\pi_1 \in \{0.5, 0.65, 0.8\}$ (across the columns). 
At each $\pi_1$ value, in line with the size results from \Cref{fig:level_prop}, the power of DP-TOST approaches that of the non-private benchmark as $n$ and/or $\epsilon$ increases.
Across the considered sample sizes, moderate or large privacy budgets ($\epsilon \geq 0.5$) lead to a statistical power for DP-TOST that closely resembles that of the non-private TOST. For smaller $\epsilon$ values, the power of DP-TOST declines quickly at smaller sample sizes ($n\leq 400$), reflecting the expected trade-off between privacy and power, due to the additive zero-mean noise used to ensure $\epsilon$-DP, which increases the variance of the test statistic without altering its expectation. 
Consequently, the DP-TOST leads to a level-$\alpha$ test procedure, even at small sample sizes, and satisfies $\epsilon$-DP at the cost of a loss in statistical power compared to its non-private counterpart that is controlled by the privacy budget $\epsilon$.

\FloatBarrier
\subsection{Bounded Means}

\begin{figure}[h]
    \centering
    \includegraphics[width=0.72\textwidth]{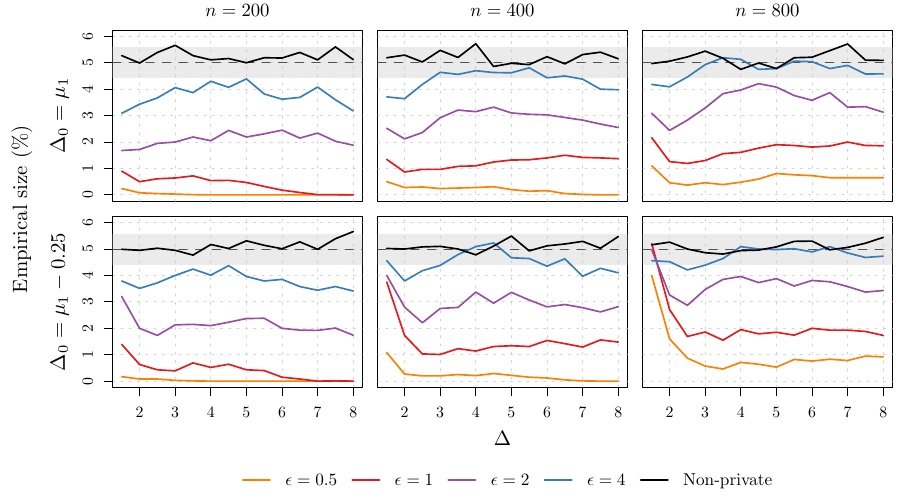}
    \caption{Empirical test size (y-axis) as a function of the half-width $\Delta$ used to clamp the data (x-axis) for the non-private TOST and DP-TOST with different privacy budgets $\epsilon$ when assessing equivalence for the difference of two means. Each row corresponds to a different data clamping scheme, and each column represents a given sample size $n \in \{200, 400, 800\}$. The gray shaded region denotes the 95\% simulation error around the nominal significance level $\alpha=5\%$.}
    \label{fig:level_means}
\end{figure}

We then evaluate the performance of the DP-TOST framework for the equivalence of means using $\sigma_1=\sigma_2=1$ and $c_0=0.5$.
Unlike proportion data that are naturally bounded, this setting requires clamping the data to bound the global sensitivity of summary statistics as required for DP. 
To assess type-I error rates, we consider two clamping schemes, where we fix the center $\Delta_0 \in \{\mu_1-0.25, \mu_1 \}$, and clamp all observations that do not belong to the interval $[a, b]$, with $a=\Delta_0 - \Delta$ and $b=\Delta_0 + \Delta$, in each of the two samples.
\Cref{fig:level_means} shows the empirical test size (y-axis), corresponding to the maximum probability of rejecting H$_0$ when $ \mu_1 - \mu_2 \in \{- c_0, c_0$\}, as a function of clamping half-widths $\Delta \in [1.5, 8]$ (x-axis), for increasing sample sizes $n$ (columns) and clamping centers $\Delta_0$ (rows).
The simulation results indicate that DP-TOST maintains type-I error control across privacy budgets $\epsilon$ and clamping half-widths $\Delta$. 
While smaller $\Delta$ values reduce global sensitivity, they lead to a higher risk of overly compressing the signal. Conversely, as $\Delta$ increases, the signal is preserved but the data are subject to higher levels of DP-noise. 
Although the empirical test size remains stable under symmetric data clamping around $\mu_1$, too narrow and asymmetric clamping can induce instability under stricter privacy budgets $\epsilon$, even while maintaining type-I error control.
This highlights the critical role of the data curator in avoiding overly restrictive bounds, as this choice directly affects the performance of the DP-TOST procedure.

\begin{figure}[t]
    \centering
    \includegraphics[width=0.85\textwidth]{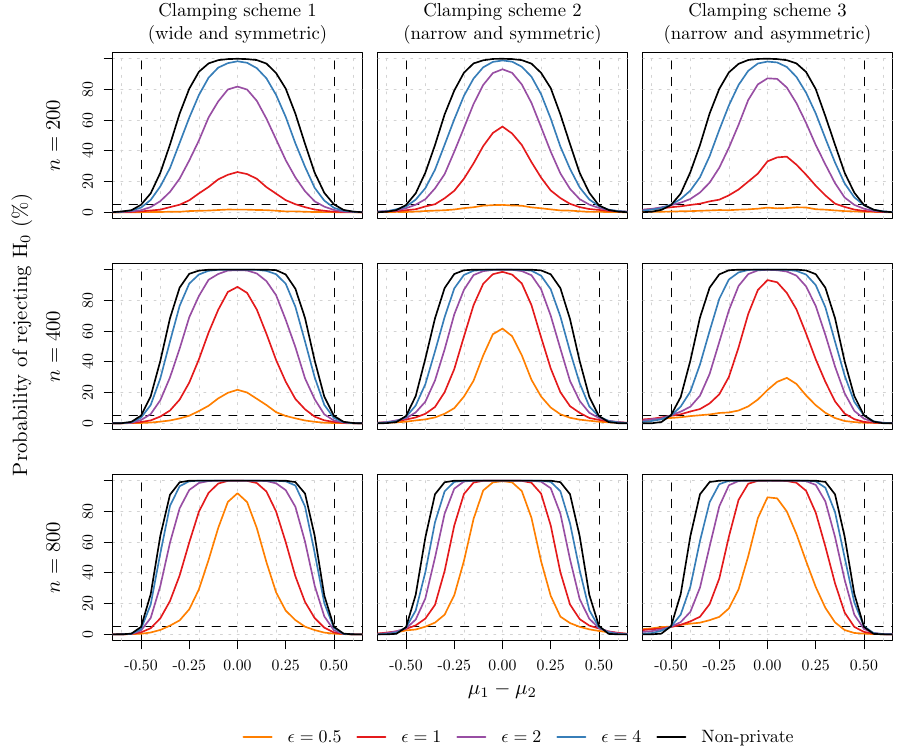}
    \caption{Empirical probability of rejecting H$_0$ (y-axis) as a function of the true difference in means (x-axis) for the non-private TOST and DP-TOST with different privacy budgets $\epsilon$. Each column corresponds to a different data clamping scheme, and each row represents a given sample size $n \in \{ 200, 400, 800\}$. Vertical dashed lines denote the equivalence margins $c_0=0.5$, and the horizontal dashed line represents the nominal significance level $\alpha=5\%$.}
    \label{fig:power_means}
\end{figure}

\Cref{fig:power_means} shows the probability of rejecting H$_0$ (y-axis) as a function of the true difference in means $\mu_1-\mu_2$ (x-axis), across increasing sample sizes $n$ (rows) and three different clamping schemes (columns). 
These schemes correspond to: 
(i) wide and symmetric clamping with $\Delta_0=\mu_1$ and $\Delta = 2\sigma_1$; 
(ii) narrow and symmetric clamping with $\Delta_0=\mu_1$ and $\Delta = 1.15\sigma_1$;
(iii) narrow and asymmetric clamping based on the interval $[\mu_1-0.84 \sigma_1, \mu_1 + 1.3\sigma_1]$.
We remark that tighter clamping margins reduce sensitivity and consequent DP noise, which typically leads to an improved statistical power when the bounds are symmetric.
However, asymmetric clamping can break the symmetry of the power function (see third column). Results for separate clamping schemes on the two samples lead to similar patterns and are omitted for brevity.
In line with the results for proportions, the power of DP-TOST increases with larger sample sizes and/or privacy budgets. At $\epsilon\geq 2$, the DP-TOST procedure achieves power comparable to the non-private TOST even with moderate sample sizes (e.g. $n=200$). For stricter privacy budgets ($\epsilon \leq 1$), the DP noise added to both the sample mean and standard deviation inflates the variance of the test statistic, resulting in a flattening of the corresponding power curves. Consequently, larger sample sizes are required to maintain appropriate power when $\epsilon$ is low. 
Overall, the simulations confirm that DP-TOST provides a valid albeit at times conservative test for mean equivalence, with performance converging to the non-private TOST as the privacy budget $\epsilon$ becomes less restrictive or the sample size $n$ grows.

\FloatBarrier
\section{Case Study}\label{sec6}

We illustrate the use of the proposed DP-TOST framework on real-world applications using the publicly available ``ACTG175'' dataset. This refers to an HIV-1 clinical trial consisting of a double-blind study to evaluate different treatments in infected adults whose CD4 cell counts range from 200 to 500 per cubic millimeter \citep{hammer1996trial}. 
The dataset, obtained from the \texttt{speff2trial} package in \texttt{R} \citep{speff2trial}, encompasses 2,139 observations across 27 variables, including demographic characteristics, baseline clinical markers (e.g. CD4 counts), and treatment outcomes collected from four randomized arms comparing different antiretroviral therapies. 
The four treatment arms encompass: (i) 600 mg of zidovudine alone (ZDV); (ii) 600 mg of zidovudine plus 400 mg of didanosine (ZDV+ddI); (iii) 600 mg of zidovudine plus 2.25 mg of zalcitabine (ZDV+ddC); and (iv) 400 mg of didanosine alone (ddI). 

In such clinical trials, it is often of interest to assess whether the outcomes obtained by different treatments or healthcare providers are in fact equivalent.
However, patient confidentiality is of great importance in this context, as disclosing HIV status or specific clinical markers leads to significant privacy concerns. Notably, even de-identified or aggregated health data can be vulnerable to patient re-identification through auxiliary information \citep{Terry2012, Masood2018, Xia2021}. 
To tackle these problems, our DP-TOST framework allows one to assess equivalence across providers (e.g. in terms of drop-out rates) or treatments (e.g. based on CD4 cell counts) without compromising patient privacy.

\begin{figure}[h]
    \centering
    \includegraphics[width=.45\textwidth]{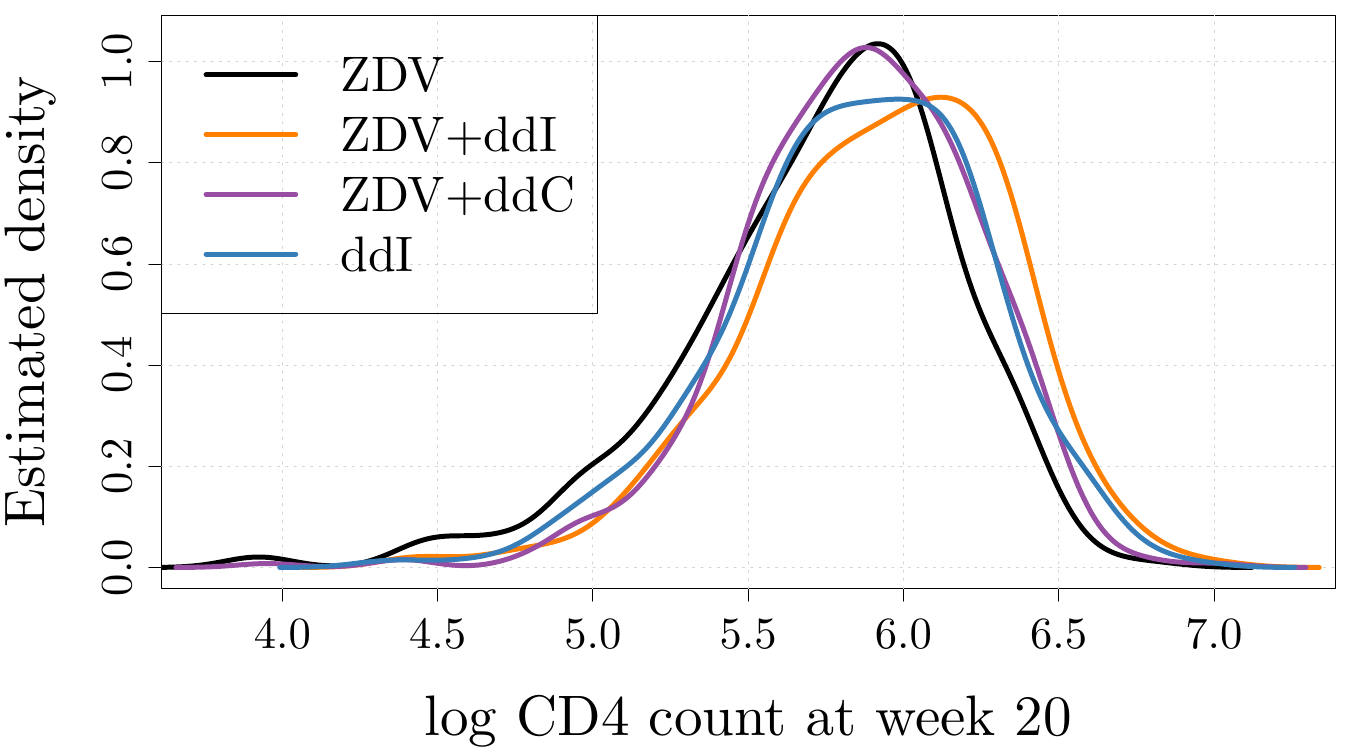}
    \caption{Kernel density estimates of log-transformed CD4 T-cell counts at week 20 (CD4\textsubscript{20}) for the four treatment arms: zidovudine alone (ZDV), zidovudine plus didanosine (ZDV+ddI), zidovudine plus zalcitabine (ZDV+ddC), and didanosine alone (ddI).}
    \label{fig:caseden}
\end{figure}

For our empirical evaluation we focus on two outcomes: (i) the proportion of participants who discontinued treatment before 96 weeks (denoted as ``Off-Treat''); and (ii) CD4 T-cell count at $20 \pm 5$ weeks (denoted as CD4\textsubscript{20}) after a natural logarithmic transformation. 
Each of these two outcomes is measured across the four treatment arms (ZDV, ZDV+ddI, ZDV+ddC, and ddI), and the resulting summary statistics are provided in \Cref{tab:cd4_summary}. The logarithm of CD4\textsubscript{20} is used to better approximate normality, as illustrated in \Cref{fig:caseden}. 

\begin{table}[h]
\centering
\caption{Descriptive statistics for the ACTG175 clinical trial data in terms of sample size, mean and standard deviation (SD) for log-transformed CD4 counts at week 20 (CD4\textsubscript{20}), and the proportion of patients discontinuing treatment before week 96 (Off-Treat) across the four treatment arms.}
\label{tab:cd4_summary}
\begin{tabular}{lcccc}
\toprule
\textbf{Treatment Arm} & \textbf{Sample Size} & \textbf{Mean log(CD4\textsubscript{20})} & \textbf{SD log(CD4\textsubscript{20})} & \textbf{Proportion Off-Treat} \\
\midrule
Zidovudine alone (ZDV) & 532 & 5.78 & 0.34 & 0.41 \\
Zidovudine plus didanosine (ZDV+ddI)               & 522 & 5.94 & 0.37 & 0.33 \\
Zidovudine plus zalcitabine (ZDV+ddC) & 524 & 5.87 & 0.34 & 0.39 \\
Didanosine alone (ddI) & 561 & 5.87 & 0.36 & 0.33 \\
\bottomrule
\end{tabular}
\end{table}

In a typical application, a data curator would release a single privatized statistic at a chosen privacy budget. However, to more accurately evaluate the operating characteristics of DP-TOST, we present the results for a Monte Carlo emulation study. Namely, for each outcome of interest, we use the summary statistics in \Cref{tab:cd4_summary} as population parameters, and generate $B=1,000$ independent synthetic replicates of the trial.
For each replicate, we perform all pairwise comparisons between the four treatment arms (resulting in 6 comparisons per outcome), and all tests are conducted at a significance level of $\alpha = 5\%$.
The standard non-private TOST is used as a benchmark to assess whether a given realization leads to a declaration of equivalence. Then, such synthetic data are privatized by adding DP noise according to a privacy budget $\epsilon$, and the DP-TOST is used to obtain another assessment of equivalence.
Results are presented in terms of agreement/disagreements between the non-private TOST and DP-TOST decisions. We are particularly interested in how often the additional DP noise alters the non-private statistical decision, to investigate whether DP noise leads to a loss of power (failing to detect true equivalence) or an inflation of type I error (falsely declaring equivalence).

\begin{figure}[t]
\centering
\includegraphics[width=0.95\textwidth]{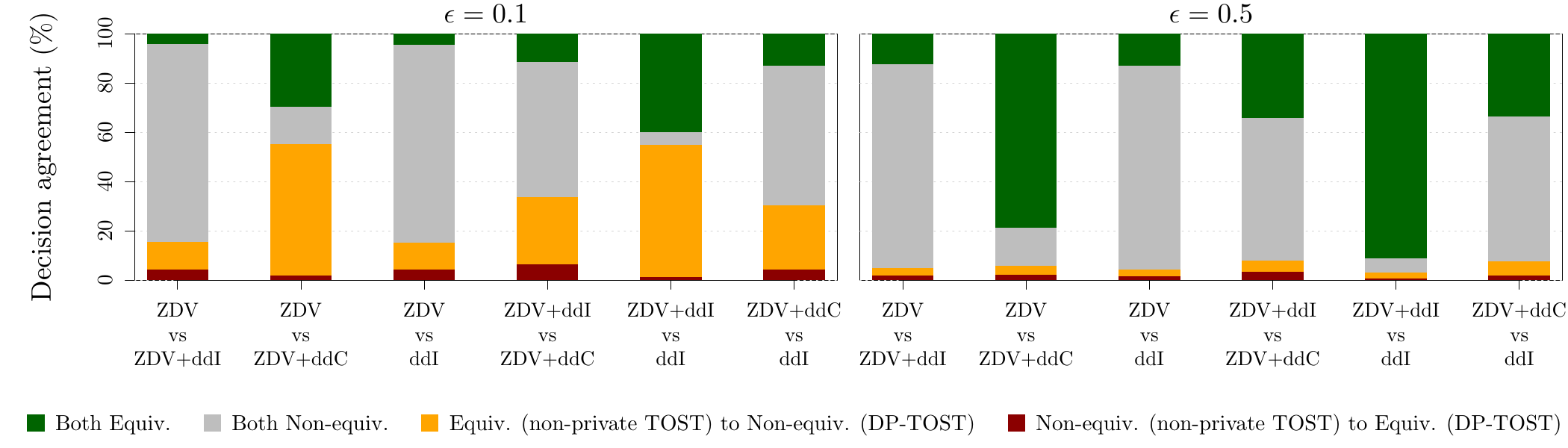}
\caption{Decision agreement for equivalence testing of "Off-Treat" proportions between the non-private TOST and DP-TOST across varying privacy budgets $\epsilon \in \{0.1, 0.5\}$ (from left to right panel). The x-axis represents pairwise comparisons between treatment arms. The stacked bars represent the percentage of 1,000 Monte Carlo replicates resulting in each category: both equivalent (green), both non-equivalent (grey), non-private equivalent/private non-equivalent (orange), and non-private non-equivalent/private equivalent (red).}
\label{fig:app_prop}
\end{figure}

\begin{table}[t]
\centering
\footnotesize
\caption{Emulation results for ``Off-Treat''  proportions comparing the equivalence decisions for non-private TOST and DP-TOST when $\epsilon=1$. Values represent the percentages across 1,000 replicates. ``Discordance'' between the two methods highlights the rate of power loss (DP-TOST cannot reject H$_0$) versus potential type I error inflation (DP-TOST rejects H$_0$).}
\label{tab:app_prop}
\begin{tabular}{
  l                       
  S[table-format=-1.2]    
  c                       
  c                       
  S[table-format=3.1]     
  S[table-format=3.1]     
  S[table-format=3.1]     
  S[table-format=3.1]     
  S[table-format=2.1]     
  S[table-format=2.1]     
}
  \toprule
  & & & & \multicolumn{2}{c}{\textbf{Non-private TOST}} & \multicolumn{2}{c}{\textbf{DP-TOST}} & \multicolumn{2}{c}{\textbf{Discordance}} \\
  \cmidrule(lr){5-6} \cmidrule(lr){7-8} \cmidrule(lr){9-10}
  
  Comparison & {$\pi_1-\pi_2$} & 
  {\makecell{Within\\$(-c_0, c_0)$?}} &
  $\epsilon$ &
  {\makecell{Cannot\\reject H$_0$}} & {\makecell{Reject\\H$_0$}} & 
  {\makecell{Cannot\\reject H$_0$}} & {\makecell{Reject\\H$_0$}} & 
  {\makecell{DP-TOST\\rejects H$_0$}} & {\makecell{DP-TOST\\cannot reject H$_0$}} \\
  \midrule

  ZDV vs ZDV+ddI & 0.08 & {\color{green!70!black}\cmark} & 0.1 & 84.3 & 15.7 & 91.7 & 8.3 & 4.1 & 11.5 \\ 
  ZDV vs ZDV+ddC & 0.02 & {\color{green!70!black}\cmark} & 0.1 & 17.2 & 82.8 & 68.5 & 31.5 & 1.9 & 53.2 \\ 
  ZDV vs ddI & 0.08 & {\color{green!70!black}\cmark} & 0.1 & 84.4 & 15.6 & 91.1 & 8.9 & 4.2 & 10.9 \\ 
  ZDV+ddI vs ZDV+ddC & -0.06 & {\color{green!70!black}\cmark} & 0.1 & 61.2 & 38.8 & 82.1 & 17.9 & 6.3 & 27.2 \\ 
  ZDV+ddI vs ddI & 0 & {\color{green!70!black}\cmark} & 0.1 & 6.3 & 93.7 & 59 & 41 & 1.1 & 53.8 \\ 
  ZDV+ddC vs ddI & 0.06 & {\color{green!70!black}\cmark} & 0.1 & 60.9 & 39.1 & 82.8 & 17.2 & 4.2 & 26.1 \\ 
  \midrule 
  ZDV vs ZDV+ddI & 0.08 & {\color{green!70!black}\cmark} & 0.5 & 84.3 & 15.7 & 85.8 & 14.2 & 1.7 & 3.2 \\ 
  ZDV vs ZDV+ddC & 0.02 & {\color{green!70!black}\cmark} & 0.5 & 17.2 & 82.8 & 19.1 & 80.9 & 2 & 3.9 \\ 
  ZDV vs ddI & 0.08 & {\color{green!70!black}\cmark} & 0.5 & 84.4 & 15.6 & 85.5 & 14.5 & 1.6 & 2.7 \\ 
  ZDV+ddI vs ZDV+ddC & -0.06 & {\color{green!70!black}\cmark} & 0.5 & 61.2 & 38.8 & 62.3 & 37.7 & 3.4 & 4.5 \\ 
  ZDV+ddI vs ddI & 0 & {\color{green!70!black}\cmark} & 0.5 & 6.3 & 93.7 & 8 & 92 & 0.7 & 2.4 \\ 
  ZDV+ddC vs ddI & 0.06 & {\color{green!70!black}\cmark} & 0.5 & 60.9 & 39.1 & 64.6 & 35.4 & 1.9 & 5.6 \\ 
  \bottomrule
\end{tabular}
\end{table}

For the proportion endpoints based on discontinuation rates (Off-Treat), we use an equivalence margin of $c_0 = 0.1$ and privacy budgets $\epsilon \in \{0.1, 0.5\}$.
\Cref{fig:app_prop} presents the results for this emulation study, and \Cref{tab:app_prop} shows a detailed breakdown.
Here, as all true effects $\pi_1-\pi_2$ belong to the alternative hypothesis space, these results highlight the trade-off between DP and test power. In treatment arms comparisons where the non-private TOST tends to conclude equivalence more often (e.g. ZDV+ddI vs. ddI, where $\pi_1 - \pi_2 \approx 0$), the DP-TOST shows high agreement rates even at moderate privacy budgets ($\epsilon \geq 0.5$). However, at strict privacy levels ($\epsilon = 0.1$), the injection of DP noise frequently leads to a loss in power, where the non-private TOST declares equivalence, but the DP-TOST cannot do so due to the inflated noise. Similar patterns hold for the comparison of other treatment arms, and we observe that DP-TOST closely matches the decision of non-private TOST when the latter does not declare equivalence, even under very strict privacy budgets.
Importantly, there are only very sporadic occurrences (especially under stricter privacy budgets) where DP-TOST declares equivalence when the non-private TOST does not, ensuring the procedure remains conservative and safe in terms of patients' risk.

\begin{figure}[t]
\centering
\includegraphics[width=1\textwidth]{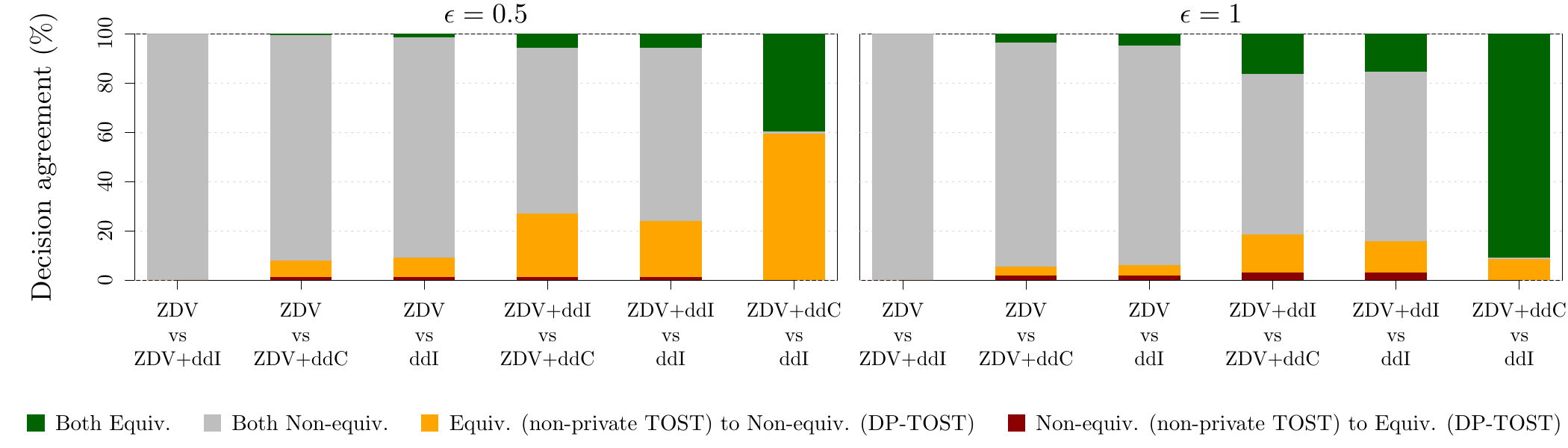}
\caption{Decision agreement for equivalence testing of mean log-transformed CD4 counts (CD4\textsubscript{20}) between the non-private TOST and DP-TOST across varying privacy budgets $\epsilon \in \{0.5, 1\}$ (from left to right panel). The x-axis represents pairwise comparisons between treatment arms. The stacked bars represent the percentage of 1,000 Monte Carlo replicates resulting in each category: both equivalent (green), both non-equivalent (grey), non-private equivalent/private non-equivalent (orange), and non-private non-equivalent/private equivalent (red).}
\label{fig:app_mean}
\end{figure}

We next present the results for the analysis of mean CD4 T-cell counts at week 20, where we consider an equivalence margin of $c_0 = \log(1.10)$ and privacy budgets $\epsilon \in \{ 0.5, 1\}$. Data are clamped within clinically plausible bounds of $[\log(100), \log(1500)]$, resulting in approximately 1-3\% clamping (see e.g. Battistini et al.\citep{BattistiniGarcia2025}). 
The results for this emulation scenario are provided in \Cref{fig:app_mean}, and \Cref{tab:app_means} presents a detailed breakdown. Overall, these results align with the ones for proportions. They indicate that even under strict privacy budgets ($\epsilon=0.5$), the non-private TOST and DP-TOST have a high level of agreement (greater than 60\%) when the former tends not to declare equivalence (i.e. all but ZDV+ddC vs ddI comparisons). The comparison between ZDV+ddC vs ddl behaves differently since the non-private TOST almost always leads to a declaration of equivalence, while DP-TOST achieves so only under larger privacy budgets ($\epsilon \geq 1$). Therefore, moderate or large privacy budgets are required for DP-TOST to achieve high levels of agreements (occurring in more than 90\% of the instances when $\epsilon = 1$).

\begin{table}[t]
\centering
\footnotesize
\caption{Emulation results for ``Off-Treat''  proportions comparing the equivalence decisions for non-private TOST and DP-TOST when $\epsilon=1$. Values represent the percentages across 1,000 replicates. ``Discordance'' between the two methods highlights the rate of power loss (DP-TOST cannot reject H$_0$) versus potential type I error inflation (DP-TOST rejects H$_0$).}
\label{tab:app_means}
\begin{tabular}{
  l                       
  S[table-format=-1.2]    
  c                       
  c                       
  S[table-format=3.1]     
  S[table-format=3.1]     
  S[table-format=3.1]     
  S[table-format=3.1]     
  S[table-format=2.1]     
  S[table-format=2.1]     
}
  \toprule
  & & & & \multicolumn{2}{c}{\textbf{Non-private TOST}} & \multicolumn{2}{c}{\textbf{DP-TOST}} & \multicolumn{2}{c}{\textbf{Discordance}} \\
  \cmidrule(lr){5-6} \cmidrule(lr){7-8} \cmidrule(lr){9-10}
  
  Comparison & {$\mu_1-\mu_2$} & 
  {\makecell{Within\\$(-c_0, c_0)$?}} &
  $\epsilon$ &
  {\makecell{Cannot\\reject H$_0$}} & {\makecell{Reject\\H$_0$}} & 
  {\makecell{Cannot\\reject H$_0$}} & {\makecell{Reject\\H$_0$}} & 
  {\makecell{DP-TOST\\rejects H$_0$}} & {\makecell{DP-TOST\\cannot reject H$_0$}} \\
  \midrule

  ZDV vs ZDV+ddI & -0.16 & {\color{red!90!black}\xmark} & 0.5 & 100.00 & 0.00 & 100.00 & 0.00 & 0.00 & 0.00 \\ 
  ZDV vs ZDV+ddC & -0.09 & {\color{green!70!black}\cmark} & 0.5 & 92.80 & 7.20 & 99.90 & 0.10 & 0.10 & 7.20 \\ 
  ZDV vs ddI & -0.09 & {\color{green!70!black}\cmark} & 0.5 & 90.70 & 9.30 & 100.00 & 0.00 & 0.00 & 9.30 \\ 
  ZDV+ddI vs ZDV+ddC & 0.07 & {\color{green!70!black}\cmark} & 0.5 & 68.10 & 31.90 & 99.90 & 0.10 & 0.00 & 31.80 \\ 
  ZDV+ddI vs ddI & 0.07 & {\color{green!70!black}\cmark} & 0.5 & 71.70 & 28.30 & 99.70 & 0.30 & 0.10 & 28.10 \\ 
  ZDV+ddC vs ddI & 0 & {\color{green!70!black}\cmark} & 0.5 & 0.70 & 99.30 & 99.10 & 0.90 & 0.00 & 98.40 \\ 
  \midrule 
  ZDV vs ZDV+ddI & -0.16 & {\color{red!90!black}\xmark} & 1 & 100.00 & 0.00 & 100.00 & 0.00 & 0.00 & 0.00 \\ 
  ZDV vs ZDV+ddC & -0.09 & {\color{green!70!black}\cmark} & 1 & 92.80 & 7.20 & 98.00 & 2.00 & 1.30 & 6.50 \\ 
  ZDV vs ddI & -0.09 & {\color{green!70!black}\cmark} & 1 & 90.70 & 9.30 & 97.30 & 2.70 & 1.20 & 7.80 \\ 
  ZDV+ddI vs ZDV+ddC & 0.07 & {\color{green!70!black}\cmark} & 1 & 68.10 & 31.90 & 93.00 & 7.00 & 1.10 & 26.00 \\ 
  ZDV+ddI vs ddI & 0.07 & {\color{green!70!black}\cmark} & 1 & 71.70 & 28.30 & 93.00 & 7.00 & 1.30 & 22.60 \\ 
  ZDV+ddC vs ddI & 0 & {\color{green!70!black}\cmark} & 1 & 0.70 & 99.30 & 60.20 & 39.80 & 0.00 & 59.50 \\ 
  \bottomrule
\end{tabular}
\end{table}

Overall, the results of this emulation study are in line with the simulation results presented in \Cref{sec5}, and they confirm that the DP-TOST framework offers privacy guarantees at the cost of a more conservative behavior compared to the non-private TOST. Indeed, although the two procedures behave comparably under large privacy budgets, for stricter budgets their discordance almost exclusively leads to DP-TOST failing to reject H$_0$ when the non-private TOST does so, indicating that DP-TOST does not lead to an inflation of type I error rates.
This is an important property in equivalence testing, ensuring that privacy protection does not lead to a higher rate of unjustified claims of equivalence.
Our findings suggest that for datasets of similar size ($n \approx 500$ per arm), privacy budgets of $\epsilon \approx 0.5$ for proportions and $\epsilon \approx 1$ for means assessments provide an effective balance between DP guarantees offered by DP-TOST and the preservation of statistical power compared to its non-private counterpart.

\FloatBarrier
\section{Final Remarks}\label{sec7}

This paper introduces a unified framework for conducting equivalence testing under formal privacy guarantees. We develop DP analogues of the widely used TOST procedure for both differences in binomial proportions and normal means by constructing confidence intervals via simulation-based matching methods. Our approach offers statistical inference while adhering to $\epsilon$-DP constraints, a critical requirement in modern biomedical and healthcare applications \citep{dwork2014algorithmic,liu2023survey}. Through an extensive simulation study, we demonstrate that the proposed DP-TOST framework maintains type-I error control and generally achieves competitive power compared to its non-private counterparts. However, as expected in DP scenarios, DP-TOST may yield overly conservative test procedures for smaller sample sizes and stricter privacy budgets. The methods remain reasonably robust across a variety of sample sizes and privacy levels, and perform comparably even in scenarios involving asymmetric clamping or group-specific bounds. Our real-data application to the ACTG175 HIV clinical trial illustrates the practical utility of the method. Equivalence conclusions obtained under DP-TOST generally mirror those from classical TOST, confirming the method's reliability in practice for the considered settings. The observed widening of confidence intervals under stronger privacy constraints (e.g. $\epsilon = 0.5, 1$) is expected but does not appear to considerably alter inference in most clinically relevant comparisons. Overall, these results indicate that DP-TOST provides the required balance between privacy and statistical validity, supporting its use in privacy-sensitive applications, such as comparing response or event rates in clinical trials. 

This work lays the foundations for extending equivalence testing under differential privacy to more complex and clinically relevant settings. Indeed, the simulation-based nature of the proposed approach allows it to be adaptable to more general statistical procedures. These include procedures addressing outcomes such as ordinal or skewed clinical endpoints, longitudinal data structures with repeated measures over time, and multi-arm trial designs commonly encountered in comparative effectiveness research. Future work may also incorporate adaptive privacy mechanisms that balance utility and confidentiality based on data sensitivity, as well as alternative test statistics that go beyond mean-based summaries, such as rank-based or survival-based measures, to accommodate diverse outcome types. Multivariate extensions may further enable applications in settings where equivalence testing is routinely applied, including bioequivalence, safety monitoring, and health outcomes research \citep{hauschke2007bioequivalence,hoffelder2015multivariate,pallmann2017simultaneous}. Overall, these developments highlight the feasibility of conducting statistically rigorous and privacy-preserving equivalence analyses in biomedical research, where protection of individual-level data remains both an ethical obligation and a regulatory requirement.
\bibliographystyle{unsrtnat}
\bibliography{references}

\end{document}